# QUANTITATIVE MAGNETIC RESONANCE IMAGE ANALYSIS VIA THE EM ALGORITHM WITH STOCHASTIC VARIATION[1]

BY XIAOXI ZHANG, TIMOTHY D. JOHNSON, RODERICK J. A. LITTLE
AND YUE CAO

*University of Michigan*

Quantitative Magnetic Resonance Imaging (qMRI) provides researchers insight into pathological and physiological alterations of living tissue, with the help of which researchers hope to predict (local) therapeutic efficacy early and determine optimal treatment schedule. However, the analysis of qMRI has been limited to ad-hoc heuristic methods. Our research provides a powerful statistical framework for image analysis and sheds light on future localized adaptive treatment regimes tailored to the individual's response. We assume in an imperfect world we only observe a blurred and noisy version of the underlying pathological/physiological changes via qMRI, due to measurement errors or unpredictable influences. We use a hidden Markov random field to model the spatial dependence in the data and develop a maximum likelihood approach via the Expectation–Maximization algorithm with stochastic variation. An important improvement over previous work is the assessment of variability in parameter estimation, which is the valid basis for statistical inference. More importantly, we focus on the expected changes rather than image segmentation. Our research has shown that the approach is powerful in both simulation studies and on a real dataset, while quite robust in the presence of some model assumption violations.

**1. Introduction.** Quantitative Magnetic Resonance Imaging (qMRI) is a noninvasive tool for visualizing the inside of living organisms, and is used to assess pathological and physiological alterations in living tissue, such as the brain. More recently, it has been used to measure physiological changes (such as diffusion, perfusion, vascular permeability and metabolism) in diseased tissue due to therapy [e.g., Cao et al. (2005), Moffat et al. (2005), Hamstra et al. (2005)]. With the aid of qMRI, investigators hope to predict

Received June 2007; revised December 2007.
[1]Supported in part by NIH Grants PO1 CA087684-5 and PO1 CA59827-11A1.

(local) therapeutic efficacy early during treatment so that treatments can be tailored to the individual.

This work is motivated by a pilot qMRI study conducted at the University of Michigan School of Medicine. Eleven volunteers with primary, high-grade gliomas were recruited for the study. Prior to the initiation of radiation therapy, the volunteers underwent T1-weighted qMRI with and without contrast enhancement. The same imaging protocol was subsequently performed after approximately the first and third week of radiotherapy, and 1, 3 and 6 months after completion of radiotherapy. The contrast agent used was Gadolinium diethylenetriaminepentaacetic acid (Gd-DTPA). Gd-DTPA has a molecular diameter in the range of many chemotherapeutic molecules and, hence, its uptake rate can be used as a surrogate of tumor/brain vascular permeability to these drugs [Cao et al. (2005)]. This was the first study to use quantitative and high-resolution MRI to assess the effects of radiation on the vascular permeability in tumors and healthy tissue to a molecule in the size range of chemotherapeutic agents in high-grade gliomas [Cao et al. (2005)].

Prognosis for primary high-grade gliomas is poor and advances in radiotherapy followed by chemotherapy have failed to prolong the median survival time of these patients beyond about 1 year after diagnosis. Chemotherapy has been largely unsuccessful in part due to the tight endothelial junctions in the tumor (blood–tumor barrier, BTB) that limit the delivery of these large chemotherapeutic molecules to the tumor cells. One of the goals of this study was to determine the effects, over time, of radiation therapy on the BTB relative to the blood–brain barrier (BBB). If it can be demonstrated that radiation therapy transiently increases the vascular permeability of the tumor to these large chemotherapeutic drugs, this may suggest an optimal time for delivering these drugs during radiation therapy, as opposed to waiting for the completion of radiation therapy. In this manuscript, we focus on the change in contrast uptake from the baseline imaging study to the 3-week imaging study, as this time point is of special interest to the investigators.

There is a large body of research on medical image analysis, in particular, functional MRI (fMRI). The fMRI analyses are typically of a time series nature [Worsley and Friston (1995)]. Random Field Theory (RFT) is applied after smoothing the data with an isotropic Gaussian filter. It assumes spatial continuity, which is reasonable for healthy volunteers in fMRI experiments. However, qMRI in the motivating study was performed on patients with solid mass tumors, where a sharp transition between the regions with highly leaky vasculature and with intact vessels is visible [Figures 2(a) and 4(a)]. The tumors are physiologically different from surrounding healthy tissue, and their contrast uptake is highly heterogeneous. The smoothing techniques that are applied in most image analyses blur tissue boundaries, and hence, do not model this feature well. Furthermore, many qMRI analyses ignore the inherent spatial correlation in the data (at the pixel level), and treat the data as independent observations [e.g., Cao et al. (2005), Moffat et al. (2005), Hamstra et al. (2005)], which can lead to incorrect variance estimates and invalid hypothesis tests.

In this manuscript we analyze qMRI data using a flexible, yet conceptually simple hidden Markov Random Field model [MRF, Besag (1974)] that accounts for spatial correlation and preserves boundaries. It is also known as the Potts model in the statistical physics literature [Potts (1952)]. The rationale for this model is that qMRI produces a blurred and noisy version of the underlying change in contrast uptake, due to measurement error, electronic noise, limited spatial resolution, reconstruction error, the filter used in the MRI reconstruction, field inhomogeneities and other unpredictable influences. The underlying pathological/physiological changes, which we call the "true" scene, consist of a finite number of homogeneous regions. To model the spatial dependence inherent in the data, we introduce a layer of latent discrete labels, and map the labels to changes in contrast uptake. We use a hidden MRF to model the spatial dependence of the data. We treat these unobservable MRF labels as missing data, and implement the Expectation–Maximization [EM, Dempster, Laird and Rubin (1977)] algorithm with stochastic variation [Wei and Tanner (1990)]. Our implementation of the EM algorithm maximizes the observed likelihood by integrating out the missing labels in the complete data likelihood.

A few comments regarding the choice of the above model are in order. We are not attaching any physical interpretation to the hidden labels. (However, labels in our model do correspond to relative changes in contrast uptake.) Rather, we impose a MRF on the labels to model the dependence in the data, and focus on the expected change in contrast uptake. Although some imaging techniques, such as positron emission tomography (PET) and single photon emission computed tomograph (SPECT), have well-defined, mechanistically interpretable, probabilistic models that are used to model the data (radioactive decay follows a Poisson process), there are no mechanistically-based models for MRI data. Lei and Udupa (2002) investigate the statistical properties of MRI from the physical imaging process, through the mathematical image reconstruction, to the resulting images, and conclude that the distributional assumption of normality is reasonable. Independent of their conclusion, we argue from a purely statistical perspective and call upon the central limit theorem. The MRI signal (contrast) at a single pixel is a function of the relaxation time of millions of nuclear spins for a single pulse sequence. A pulse sequence lasts on the order of a second or two, or less, and each MRI image is obtained via hundreds of signal acquisitions. Each of these signals (from each pulse sequence) is superimposed with noise: spontaneous fluctuations (errors) from various sources such as magnetic field inhomogeneities, thermal motion (Brownian motion) of free electrons in the electronics (mainly from the RF coil), motion of the imaged object, etc. Thus, it is completely reasonable to postulate Gaussian noise based on the central limit theorem when the noise is not too high [see, e.g., Liang and Lauterbur (1999), Chapter 8].

Previous implementations of the EM algorithm for the MRF model have problems or limitations. Either the likelihood was approximated or the E-step in the EM algorithm relied on an approximation, both of which can result in biased maximum likelihood estimates. In other implementations the spatial regularization parameter in the MRF was assumed to be known and results can be highly sensitive to the choice of this parameter. Melas and Wilson (2002) pointed out that the pseudo-likelihood approach tends to overestimate the regularization parameter of the MRF and over-smooth the data. Some researchers [Chalmond (1989), Won and Derin (1992), Zhang, Modestino and Lagan (1994) and Panjwani and Healey (1995)] used the pseudo-likelihood approach proposed by Besag (1974). Although Zhang, Brady and Smith (2001) and Sengur, Turkoglu and Ince (2006) correctly specified the likelihood, they made an approximation in the E-step. We show in the results section that this approximation can lead to larger misclassification rates than our proposed algorithm. Lei and Udupa (2003) used the Iterative Conditional Mode [ICM, Besag (1974, 1986)] algorithm, which they referred to as "MRF-ICM." However, they used the product of local conditional distributions rather than the actual joint distributions, which in a sense emulates Besag's pseudo-likelihood approach. These approaches maximize the complete-data likelihood jointly with respect to parameters and missing data, an approach which in general lacks the consistency and asymptotic efficiency of maximum likelihood estimates (MLE) as pointed out by Little and Rubin (1983). Furthermore, estimates of uncertainty of the parameter estimates and the hidden MRF labels are not provided [see also Deng and Clausi (2004)]. The goal of most image analyses via MRFs is segmentation, and therefore aims at a labeling of all pixels. This optimization task is often performed in an iterative fashion [Won and Derin (1992)], or via simulated annealing [Lakshmanan and Derin (1989)]. The MRF models have also found success in the fMRI literature [e.g., Descombes, Kruggel and Cramon (1998a and 1998b), Woolrich et al. (2005), Svensen, Kruggal and Cramon (2000), Rajapakse and Piyaratna (2001), Wang and Rajapakse (2006)], where segmentation remains the main goal. However, segmentation is not of primary interest in our application. In fact, the segmentation labels lack a strong biologically meaningful interpretation. Rather, we are using the hidden MRF model as a way to account for spatial correlation and as a way to perform edge-preserving smoothing of the image. It is the underlying changes that are of scientific interest.

In this manuscript we build on previous work on the hidden MRF model by (1) correctly implementing the EM algorithm with stochastic variation;

(2) estimating the spatial regularization parameter rather than assuming that it is known; (3) estimating standard errors of the parameter estimates via the Louis (1982) method; and (4) focusing on the expected local change in contrast uptake rather than segmentation.

This manuscript is organized as follows. In Section 2 we present our model and the EM algorithm, and discuss estimated standard errors and model selection. In Section 3 we present results from a simulation study where we investigate the sensitivity to model assumption violations. Results from the motivating example are presented. We conclude by summarizing the strengths and weaknesses of our approach and discussing future work.

## 2. Model and algorithm.

2.1. *Image model.* We use the following notation. Pixels (short for picture elements) will be indexed by $i = 1, 2, \ldots, N$. If pixels $i$ and $i'$ are immediately adjacent (sharing a common edge), we call them neighbors, denoted $i \sim i'$. The set of neighbors of pixel $i$ is denoted $\partial i = \{i' : i' \sim i\}$. Associated with each pixel $i$ are the observed pixel intensity $y_i$ and a hidden label $z_i$. The collection of the observed pixel intensities $\mathbf{y}^{\mathrm{T}} = (y_1, y_2, \ldots, y_N)$ is called the image (i.e., the change in contrast uptake), while the collection of latent labels $\mathbf{z}^{\mathrm{T}} = (Z_1 = z_1, Z_2 = z_2, \ldots, Z_N = z_N)$ defined on a finite discrete state space is called a configuration. The set of pixels with the same hidden label is referred to as a component, which can consist of disjoint clusters of pixels.

We assume there is an $M$-state MRF on the state space $\mathcal{S} = \{1, 2, \ldots, M\}$. Each state is mapped to an intensity in the "true" scene. It follows that there are $M^N$ configurations on the configuration space $\mathcal{S}^N$, the number of which increases exponentially with the number of pixels $N$. Our image model is a two-level hierarchical model. The higher level specifies the spatial structure of the MRF with probability mass function

$$\Pr(\mathbf{Z} = \mathbf{z} \mid \beta) = g^{-1}(\beta) \exp\left\{\sum_{i \sim i'} \beta \, \mathrm{I}(z_i = z_{i'})\right\} \qquad \text{for all } \mathbf{z} \in \mathcal{S}^N,$$

where $\mathrm{I}(\cdot)$ is the indicator function and the regularization parameter $\beta \geq 0$ controls the spatial smoothness of the MRF. The normalizing constant $g(\beta) = \sum_{\mathbf{z} \in \mathcal{S}^N} \exp\{\sum_{i \sim i'} \beta \mathrm{I}(z_i = z_{i'})\}$ has $M^N$ summands, and is not analytically tractable. Given $\mathbf{Z} = \mathbf{z}$, the observed pixel intensities are conditionally independent with Gaussian noise on the lower level,

$$y_i \mid z_i = \mu_{z_i} + e_i, \qquad e_i \sim \mathrm{N}(0, \sigma_{z_i}^2) \qquad \text{for all } i,$$

where $\mu_{z_i} = \mu_k$ and $\sigma_{z_i}^2 = \sigma_k^2$, when $z_i = k$ ($1 \leq k \leq M$). We also write $\boldsymbol{\mu} = (\mu_1, \ldots, \mu_M)^{\mathrm{T}}$, $\boldsymbol{\sigma}^2 = (\sigma_1^2, \ldots, \sigma_M^2)^{\mathrm{T}}$ and $\boldsymbol{\theta}^{\mathrm{T}} = (\boldsymbol{\mu}^{\mathrm{T}}, \boldsymbol{\sigma}^{2\mathrm{T}}, \beta)$.

A few comments are in order: (1) The model requires little prior knowledge about the spatial structure of the hidden configuration, only that neighboring pixels tend to share the same label. (2) The regularization parameter, $\beta$, controls the strength of the association between neighbors. When $\beta$ is large, the correlation between pixels is strong (neighboring pixels have high tendency to assume the same label), and the configuration tends to be smooth. Note that when $\beta = 0$ our model degenerates to a nonspatial Gaussian mixture model with equal component weights, in which case pixels are independent. The spatial correlation decreases as the distance between pixels increases. (3) Although the likelihood assumes conditionally independent Gaussian noise given the hidden labels, the data are marginally dependent. (4) It is conceptually the same to apply this model to both two dimensional and three-dimensional images. For instance, a pixel has up to four neighbors in 2D images compared to up to six in 3D images. In the 3D case, the term "voxel" is usually used instead of "pixel."

2.2. *The EM algorithm with stochastic variation.* With the introduction of the unobservable labels in the hidden MRF, the image model can be viewed as a missing data problem. The changes in contrast uptake are the observed data ($Y_{\text{obs}} = \mathbf{y}$) and the pixel labels are treated as missing data ($Y_{\text{mis}} = \mathbf{Z}$). The hidden labels are assumed to be unobserved random variables with a probability distribution.

The EM algorithm utilizes the *complete-data* log-likelihood

$$l_{\text{comp}}(\boldsymbol{\theta} \mid \mathbf{y}, \mathbf{Z} = \mathbf{z}) = -0.5 N \log(2\pi) - \sum_{k=1}^{M} \sum_{i \in D_k} \{\log(\sigma_k) + 0.5 \sigma_k^{-2}(y_i - \mu_k)^2\} + \sum_{i \sim i'} \beta \mathrm{I}(z_i = z_{i'}) - \log g(\beta)$$

to maximize the *observed* log-likelihood

$$l_{\text{obs}}(\boldsymbol{\theta} \mid \mathbf{y}) = \sum_{\mathbf{z} \in \mathcal{S}^N} l_{\text{comp}}(\boldsymbol{\theta} \mid \mathbf{y}, \mathbf{Z} = \mathbf{z}) \Pr(\mathbf{Z} = \mathbf{z} \mid \mathbf{y}, \boldsymbol{\theta}).$$

This is in contrast to most existing frequentist analyses, which treat the hidden labels as unknown but fixed and maximize the likelihood jointly with respect to the hidden labels and the model parameters.

The complete-data log-likelihood belongs to the exponential family with complete data sufficient statistics $T_{k1} = N_k$, $T_{k2} = \sum_{i \in D_k} y_i$, $T_{k3} = \sum_{i \in D_k} y_i^2$ for $k = 1, 2, \ldots, M$, and $T_4 = \sum_{i \sim i'} \mathrm{I}(z_i = z_{i'})$. The component with common label $k$ is denoted as $D_k = \{i : z_i = k\}$ with $N_k$ pixels. Hence, the EM algorithm iterates between a simplified E-step (expectation) and an M-step (maximization). At the $t$th iteration, the E-step computes the conditional expectation of the complete-data sufficient statistics given the observed data and current parameter estimates $\boldsymbol{\theta}^{(t)}$,

$$\mathbf{T}^{(t)} = \sum_{\mathbf{z} \in \mathcal{S}^N} \mathbf{T}(\mathbf{y}, \mathbf{Z} = \mathbf{z}) \Pr(\mathbf{Z} = \mathbf{z} \mid \mathbf{y}, \boldsymbol{\theta}^{(t)}), \tag{1}$$

where $\mathbf{T}(\mathbf{y}, \mathbf{Z} = \mathbf{z}) = (T_{11}, T_{12}, T_{13}, \ldots, T_{M1}, T_{M2}, T_{M3}, T_4)$, and $\Pr(\mathbf{Z} = \mathbf{z} \mid \mathbf{y}, \boldsymbol{\theta}^{(t)})$ is the conditional distribution of the latent labels. The M-step updates $\boldsymbol{\theta} = \boldsymbol{\theta}^{(t+1)}$ as the solution of the complete-data likelihood equations.

However, the conditional expectation in the E-step in our setting has $M^N$ summands, and is not analytically tractable. One solution is to approximate the expectations stochastically in the E-step, as in Monte Carlo EM (MCEM, Wei and Tanner). The $t$th EM iteration consists of the following steps:

1. E-step: Draw configurations $\mathbf{z}^{(1)}, \ldots, \mathbf{z}^{(S_t)} \sim \Pr(\mathbf{Z} \mid \mathbf{y}, \boldsymbol{\theta}^{(t)})$, and compute the Monte Carlo estimates of the conditional expectation of the sufficient statistics $\mathbf{T}(\mathbf{y}, \mathbf{z})$ given the observed data $\mathbf{y}$ and current parameter estimates $\boldsymbol{\theta}^{(t)}$,

$$T_{k1}^{(t)} = S_t^{-1} \sum_{s=1}^{S_t} N_k^{(s)}, \qquad T_{k2}^{(t)} = S_t^{-1} \sum_{s=1}^{S_t} \sum_{i \in D_k^{(s)}} y_i, \qquad T_{k3}^{(t)} = S_t^{-1} \sum_{s=1}^{S_t} \sum_{i \in D_k^{(s)}} y_i^2$$

for $k = 1, \ldots, M$,

and

$$T_4^{(t)} = S_t^{-1} \sum_{s=1}^{S_t} \sum_{i \sim i'} \mathrm{I}(z_i^{(s)} = z_{i'}^{(s)}).$$

We draw $\mathbf{z}^{(s)}$ using the Swendsen–Wang algorithm [Swendsen and Wang (1987)], an efficient sampler specifically developed for the Potts model. It updates labels for clusters of pixels rather than one pixel at a time as in ICM.

Since the complete-data log-likelihood consists of two distinct parts $l(\boldsymbol{\mu}, \boldsymbol{\sigma}^2)$ and $l(\beta)$, the M-step has two parts:

2. M1-step: update the Gaussian parameter estimates $(\boldsymbol{\mu}^{(t+1)}, \boldsymbol{\sigma}^{2(t+1)})$ based on the expectations in the E-step, $\mu_k^{(t+1)} = T_{k2}^{(t)}/T_{k1}^{(t)}$, $\sigma_k^{2(t+1)} = T_{k3}^{(t)}/T_{k1}^{(t)} - (\mu_k^{(t+1)})^2$, for $k = 1, 2, \ldots, M$.

3. M2-step: solve $l'_{\text{comp}}(\beta) = 0$ [i.e., $g'(\beta)/g(\beta) = T_4^{(t)}$] for the regularization parameter estimate $\beta^{(t+1)}$. As the monotonicity of the ratio $g'(\beta)/g(\beta)$ can be shown, any root finding algorithm can be used.

2.3. *Standard errors of parameter estimates.* We use the method of Louis to derive the asymptotic covariance matrix of the parameter estimates based on the complete loglikelihood. The observed information matrix is formulated as the difference of the complete-data information $I_{comp}$ and the information for the conditional distribution of missing data given the observed data $I_{mis}$, that is,

$$\begin{aligned} I_{obs}(\boldsymbol{\theta}) &= I_{comp}(\boldsymbol{\theta}) - I_{mis}(\boldsymbol{\theta}) \\ &= -E_{\boldsymbol{\theta}}\left(\frac{\partial^2 l_{comp}}{\partial \boldsymbol{\theta}\, \partial \boldsymbol{\theta}^T}\Big|\mathbf{y}\right) - E_{\boldsymbol{\theta}}\left(\frac{\partial l_{comp}}{\partial \boldsymbol{\theta}}\frac{\partial l_{comp}}{\partial \boldsymbol{\theta}^T}\Big|\mathbf{y}\right) \\ &\quad + E_{\boldsymbol{\theta}}\left(\frac{\partial l_{comp}}{\partial \boldsymbol{\theta}}\Big|\mathbf{y}\right) E_{\boldsymbol{\theta}}\left(\frac{\partial l_{comp}}{\partial \boldsymbol{\theta}^T}\Big|\mathbf{y}\right). \end{aligned}$$

It follows from the asymptotic properties of the MLE that $\hat{\boldsymbol{\theta}} - \boldsymbol{\theta} \sim N(0, I_{obs}^{-1}(\hat{\boldsymbol{\theta}}))$.

2.4. *Estimating the number of states in the Markov random field.* Since there is no clear substantive rationale for determining the number of states in the MRF, we use information criteria, such as the Akaike Information Criterion [AIC, Akaike (1973)] and the Bayesian Information Criterion [BIC, Schwarz (1978)]. We run the proposed algorithm for a range of values of $M$, and compute $-2l_{obs}(\hat{\boldsymbol{\theta}}_M \,|\, \mathbf{y}, M) + c(2M+1)$ based on observed log-likelihood $l_{obs}(\hat{\boldsymbol{\theta}}_M \,|\, \mathbf{y}, M)$ of an $M$-state MRF, where $c = 2$ for AIC and $c = \log N$ for BIC. Smaller AIC or BIC is preferred—there does not appear to be a consensus choice between these criteria, but in our application they lead to the same value of $M$. We use proper multiple imputation [Rubin (1987)] to approximate $l_{obs}(\hat{\boldsymbol{\theta}}_M \,|\, \mathbf{y}, M)$, via the expression

$$\hat{l}_{obs}(\hat{\boldsymbol{\theta}}_M \,|\, \mathbf{y}, M) \approx D^{-1} \sum_{d=1}^{D} l_{comp}(\hat{\boldsymbol{\theta}}_M \,|\, \mathbf{y}, \mathbf{z}_M^{(d)}, M),$$

where $\boldsymbol{\theta}_M^{(d)}$ $(d = 1, 2, \ldots, D)$ are drawn from the asymptotic distribution $N(\hat{\boldsymbol{\theta}}_M, I_{obs}^{-1}(\hat{\boldsymbol{\theta}}_M))$ and $\mathbf{z}^{(d)}$ are drawn from $\Pr(\mathbf{Z}_M \,|\, \boldsymbol{\theta}_M^{(d)}, \mathbf{y}, M)$.

2.5. *Expected change in contrast uptake.* As stated in the Introduction, the scientific interest is in the underlying change in contrast uptake, rather than the latent labels, which lack a physical interpretation. As summarized by Cappe, Mouline and Ryden (2005), this is the situation in which the labels are "completely fictitious" and the probabilistic structure of the hidden labels is "used only as a tool for modeling dependence in the data."

For estimation purposes, we integrate out the hidden labels and define

$$\mu_{z_i}^{est} = E(\mu_{Z_i} \,|\, \mathbf{y}, \boldsymbol{\theta}) = \sum_{k=1}^{M} \mu_{Z_i} \Pr(Z_i \,|\, \mathbf{y}, \boldsymbol{\theta}) = \sum_{k=1}^{M} \mu_k \Pr(Z_i = k \,|\, \mathbf{y}, \boldsymbol{\theta})$$

as the expected change of contrast uptake of pixel $i$. This estimate, $\mu_{z_i}^{\text{est}}$, is the "denoised" change in contrast uptake. In the following section, we draw samples of the hidden configuration $\mathbf{z}^{(s)}$ $(s = 1, 2, \ldots, S)$ from $\Pr(\mathbf{Z} \mid \mathbf{y}, \hat{\boldsymbol{\theta}})$, where $\hat{\boldsymbol{\theta}}$ is the MLE of $\boldsymbol{\theta}$. The Monte Carlo approximation of $\mu_{z_i}^{\text{est}}$ is $\hat{\mu}_{z_i}^{\text{est}} = S^{-1} \sum_{s=1}^{S} \hat{\mu}_{z_i^{(s)}}$, where $\hat{\mu}_{z_i^{(s)}} = \hat{\mu}_k$ when $z_i^{(s)} = k$ in the $s$th configuration $\mathbf{z}^{(s)}$.

Similarly, we define $(\sigma_{z_i}^{\text{est}})^2 = \text{Var}(\mu_{Z_i} \mid \mathbf{y}, \boldsymbol{\theta})$ as a measure of the uncertainty in the expected change in contrast uptake, and estimate it by $(\hat{\sigma}_{z_i}^{\text{est}})^2 = S^{-1} \sum_{s=1}^{S} (\hat{\mu}_{z_i^{(s)}})^2 - (\hat{\mu}_{z_i}^{\text{est}})^2$.

**3. Results.** We first conduct a simulation study under the model assumptions to evaluate the performance of the proposed method. We then apply the algorithm when the observed image is smoothed with various Gaussian smoothing kernels, to assess robustness to violations of the model assumption of conditional independence. The variance of the Gaussian kernels are specified as the spread of an (unnormalized) density at half of its maximum value (the full width at half maximum, FWHM), an approach commonly employed in signal processing. The FWHM is related to the standard deviation of a Gaussian distribution via the expression $\sigma = \text{FWHM}/(2\sqrt{2\log 2})$. A large FWHM corresponds to wide bandwidth and results in heavy smoothing and large spatial correlation. In the simulation studies, we use a superscript to denote the FWHM of the Gaussian kernel used (i.e., $\mathbf{y}^{\text{FWHM}}$), and $\mathbf{y}^0$ means no smoothing (i.e., conditionally independent noise). The results on the real dataset are also presented.

3.1. *Simulation study with conditionally independent noise.* We first simulate a pure noise image without any signal on a $128 \times 128$ lattice, that is, $y_i$'s are independently and identically distributed $N(0,1)$. A single component is selected by both information criteria. The expected change in contrast uptake is the mean of all $y_i$'s.

We then simulate a hidden configuration ($\mathbf{z}^{\text{true}}$) with ten distinct components on the same lattice, and map it into a "true" scene $\boldsymbol{\mu}_{\mathbf{z}^{\text{true}}} = (\mu_{z_1^{\text{true}}}, \mu_{z_2^{\text{true}}}, \ldots, \mu_{z_N^{\text{true}}})$ (Figure 1, top panel). We use light gray to denote high intensity. Under the conditionally independent noise assumption, we add white noise to $\boldsymbol{\mu}_{\mathbf{z}^{\text{true}}}$ to obtain the observed image, $\mathbf{y}^0$. The parameters used to generate the simulation are listed in Table 1.

We first fit a Gaussian mixture model (ignoring the spatial structure) with 10 components and equal component weights. Over half of the pixels are misclassified due to the high noise level. There is also considerable bias in the parameter estimates (Table 1). Next, we fit our proposed algorithm with $M = 6, 7, \ldots, 16$, components. The initial values of $\mu_k$, $k = 1, 2, \ldots, M$, are evenly spaced over the range of the data $(-11.2, 12.6)$. The initial values of $\sigma_k$, $k = 1, 2, \ldots, M$, are $1/(2M)$ times the range of the data. Ten components are selected by both AIC and BIC. The final estimates $\hat{\boldsymbol{\theta}}_{M=10}$ and their standard deviations are listed in Table 1. All 95% confidence intervals cover the true parameter values. The expected pixel intensity $\hat{\mu}^{\text{est}}_{z_i}$ when $M = 10$ is estimated from 500 Monte Carlo samples (Figure 1, bottom left). As a measure of the difference between $\mu^{\text{true}}_z$ and $\hat{\mu}^{\text{est}}_{z_i}$, we compute the sum of squared discrepancies, $SS_{<\text{est,true}>} = \sum_{i=1}^{N}(\hat{\mu}_{z_i^{\text{est}}} - \mu_{z_i^{\text{true}}})^2$ (smaller value suggests better fit). The sum of the squared discrepancy of the proposed algorithm is 384.32, about 2% of the sum of squares of the noise $SS_{<\text{obs}^0,\text{true}>} = \sum_{i=1}^{N}(y_i^0 - \mu^{\text{true}}_{z_i})^2 = 16539.63$.

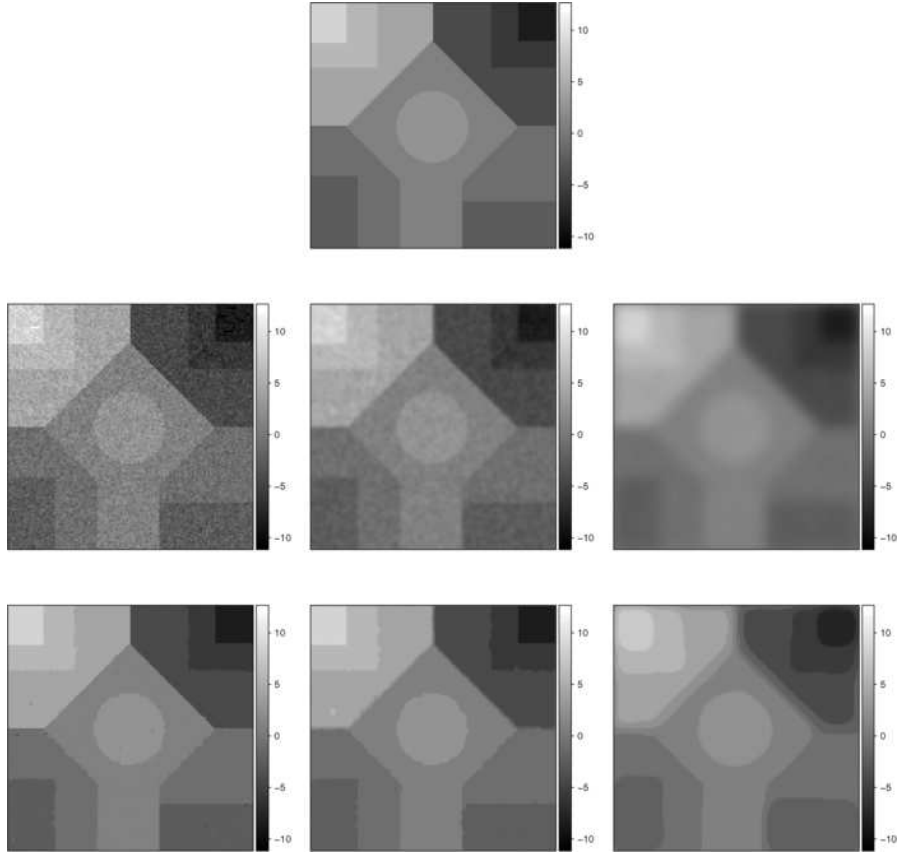

FIG. 1. *Simulation studies: the "true" scene (top), the observed images (middle), the expected pixel intensities (bottom), when* FWHM $= 0, 2$ *and* $8$ *from left to right* $(M = 10)$.

We also investigate the false positive rate (FPR) and false negative rate (FNR) under the simulation study. The proposed algorithm uniformly produces lower FPR and FNR than ignoring spatial correlation. Choosing an arbitrary threshold value of 5.0, the FPR and FNR of directly thresholding the observed image $\mathbf{y}^0$ are 3.0% and 9.8%, compared to 0.1% and 0.1% when considering the spatial structure.

TABLE 1
*True and estimated parameter values under conditionally independent noise: $\mu_k$ and $\sigma_k$ are the true parameter values; $\bar{\mu}_k$ and $\bar{\sigma}_k$ are the parameter estimates without consideration of spatial correlation; $\tilde{\mu}_k$ and $\tilde{\sigma}_k$ are parameter estimates from Zhang, Brady and Smith (2001); $\hat{\mu}_k$ and $\hat{\sigma}_k$ are parameter estimates from the proposed algorithm*

| Comp. label($k$) | Mean | | | | Standard deviation | | | |
|---|---|---|---|---|---|---|---|---|
| | $\mu_k$ | $\bar{\mu}_k$ | $\tilde{\mu}_k$ | $\hat{\mu}_k$ $(SD \times 10^2)$ | $\sigma_k$ | $\bar{\sigma}_k$ | $\tilde{\sigma}_k$ | $\hat{\sigma}_k$ $(SD \times 10^2)$ |
| 1 | $-8.50$ | $-5.93$ | $-8.57$ | $-8.56$ (4.77) | 1.00 | 2.10 | 0.94 | 0.94 (6.44) |
| 2 | $-5.95$ | $-4.38$ | $-5.98$ | $-5.93$ (3.80) | 1.00 | 0.93 | 0.97 | 1.00 (5.43) |
| 3 | $-4.25$ | $-2.80$ | $-4.31$ | $-4.28$ (2.27) | 1.00 | 0.83 | 0.97 | 0.99 (3.22) |
| 4 | $-2.55$ | $-1.50$ | $-2.56$ | $-2.54$ (2.19) | 1.00 | 0.70 | 0.97 | 1.00 (3.13) |
| 5 | $-0.85$ | $-0.54$ | $-0.85$ | $-0.85$ (1.81) | 1.00 | 0.80 | 0.97 | 0.98 (2.59) |
| 6 | 0.85 | 0.27 | 0.84 | 0.85 (1.70) | 1.00 | 0.79 | 0.99 | 1.01 (2.45) |
| 7 | 2.55 | 1.44 | 2.56 | 2.55 (3.03) | 1.00 | 0.92 | 0.95 | 0.95 (4.12) |
| 8 | 4.25 | 1.74 | 4.28 | 4.24 (2.37) | 1.00 | 1.23 | 0.98 | 1.00 (3.38) |
| 9 | 5.95 | 4.36 | 5.99 | 5.94 (3.47) | 1.00 | 1.03 | 0.93 | 0.95 (4.74) |
| 10 | 8.50 | 5.87 | 8.53 | 8.50 (5.32) | 1.00 | 2.12 | 1.02 | 1.05 (7.99) |

The expectation of the conditional loglikelihood given the observed data is

$$Q(\boldsymbol{\theta} \mid \boldsymbol{\theta}^{(t)}) = \sum_{\mathbf{z} \in \mathcal{S}^N} \Pr(\mathbf{Z} = \mathbf{z} \mid \mathbf{y}, \boldsymbol{\theta}^{(t)})\{\log f(\mathbf{y} \mid \mathbf{z}, \boldsymbol{\theta}^{(t)}) + \log \Pr(\mathbf{Z} = \mathbf{z} \mid \boldsymbol{\theta}^{(t)})\}.$$

The summation above involves $M^N$ summands and is computationally intractable. Therefore, we estimate it via Monte Carlo simulation. Zhang, Brady and Smith (2001) implement an EM-type algorithm for a similar model to ours, that we implement for comparison. In the E-step, they calculate

$$\tilde{Q}(\boldsymbol{\theta} \mid \boldsymbol{\theta}^{(t)}) = \sum_{i=1}^{N} \sum_{k=1}^{M} \Pr(Z_i = k \mid y_i, \boldsymbol{\theta}^{(t)})$$
$$\times \{\log f(y_i \mid Z_i = k, \theta^{(t)}) + \log \Pr(Z_i = k \mid \mathbf{z}_{\partial i}, \boldsymbol{\theta}^{(t)})\},$$

which is computationally tractable as there are only $M \times N$ summands. Our simulation study shows that for large images this approximation results in inferior algorithm performance, which we now discuss.

The misclassification rate (MCR) from the last configuration using the algorithm proposed by Zhang, Brady and Smith (2001) is 3.7%, compared to a MCR of 0.6% from our algorithm. The estimated configuration from their approach has many more small patches of "incorrect" labels. Moreover, their parameter estimates ($\bar{\mu}_k$ and $\bar{\sigma}_k$) generally have larger bias than our proposed method ($\hat{\mu}_k$ and $\hat{\sigma}_k$), especially in the standard deviation estimates (Table 1). We also notice that the configuration in their approach gets stuck after a few dozen iterations, which we think indicates slow mixing and a tendency to get trapped in local modes. As a matter of fact, when we initialize our algorithm with their final configuration and parameter estimates, the loglikelihood always increases. For instance, in one run it increases from $-40605$ to $-40379$.

3.2. *Simulation studies with correlated noise.* After some algebraic manipulations, one can see that, under the conditionally independent noise assumption, the observed intensities are marginally dependent [i.e., $\text{Corr}(y_i, y_{i'}) \neq 0$]. This is due to the spatial correlation induced by the MRF. However, conditional independence is still a strong assumption. Therefore, we conduct a series of simulation studies with varying degrees of correlated noise to assess the robustness of our model to violations of this model assumption.

We apply Gaussian smoothing kernels with FWHM $= 1, 2, 4$ and $8$ ($\sigma = 0.42, 0.85, 1.70$ and $3.40$) on $\mathbf{y}^0$. Due to space limitations, we display the results using FWHM $= 2$ and $8$ while fixing $M = 10$ (Figure 1). The "edge preservation" of the proposed method is evident, especially when FWHM $= 8$ (bottom row in Figure 1). However, some local features are not recovered due to extensive smoothing (e.g., the corners are rounded). Although some smoothing is intrinsic in qMRI reconstruction algorithms, the above results suggest that the common practice of smoothing before image analysis for noise-reduction purposes is not necessary when combined with our proposed method. Quantitatively, when the smoothing is relatively local (FWHM $= 1$), the estimated intensities are only slightly worse than with no smoothing ($SS_{<\text{est}1,\text{true}>} = 351.87$). When the smoothing is more global (FWHM $= 2, 4, 8$), the sum of squared discrepancies are large, $SS_{<\text{est}2,\text{true}>} = 2071.71, SS_{<\text{est}4,\text{true}>} = 4929.58$ and $SS_{<\text{est}8,\text{true}>} = 10966.29$.

3.3. *Application.* In the motivating study, eleven patients received fractionated three-dimensional conformal radiation with a median dose of 70 Gy at 2 Gy per fraction, and underwent Gd-DTPA contrast enhanced T1-weighted qMRI before, during, and after treatment. All images were registered to anatomical Computed Tomography (CT) images obtained for treatment planning purpose. The natural logarithm of the ratio of the post- and pre-enhanced T1-weighted qMRI images are used as the Gd-DTPA contrast uptake index after image normalization [Cao et al. (2005)]. We use a subset of the data, that is, the pre-radiation visit and the visit at approximately 3 weeks after the initiation. We take the change in contrast uptake from the baseline to the 3-week follow-up visit as a surrogate of the change in vascular permeability [Figures 2(a) and 4(a)], which is of special interest to the investigator. To save space, we only display the results on two patients. The other patients demonstrate similar results.

We first ignore the spatial information and pool all pixel intensities within each subject. The observed change in contrast uptake in the tumor has much heavier tails than in the healthy tissue. Although a two sample t-test suggests a statistically significant difference ($p < 0.0001$) between the tumor (mean 0.017 for subject 1 and $-0.080$ for subject 2) and the healthy tissue (mean 0.008 for subject 1 and $-0.053$ for subject 2) for both patients, the absolute difference in means is quite small—significance is driven by the extremely large number of pixels and is most likely uninteresting. More importantly, it does not provide information on the differential change in contrast uptake between the tumor and healthy tissue.

We run the proposed algorithm using several different numbers of hidden states: $M = 2, 3, \ldots, 14$. Both AIC and BIC choose $M = 3$ as the best model for subject 1. The three component mean estimates are $-0.187 \pm 0.004$ (mean $\pm$ standard deviation), $0.005 \pm 0.0005$ and $0.210 \pm 0.003$ respectively. A negative value indicates a decrease in contrast uptake, while a positive value indicates an increase. As stated earlier, we are interested in the expected change in contrast uptake, rather than the hidden labels, which may lack a biological interpretation. The expected change in contrast uptake $\hat{\mu}^{\text{est}}_{z_i}$ [Figure 2(b)] clearly shows the two concentric rings in the observed image [Figure 2(a)]. Pixels near the boundaries of components are more variable than those away from the boundaries [Figure 2(c)].

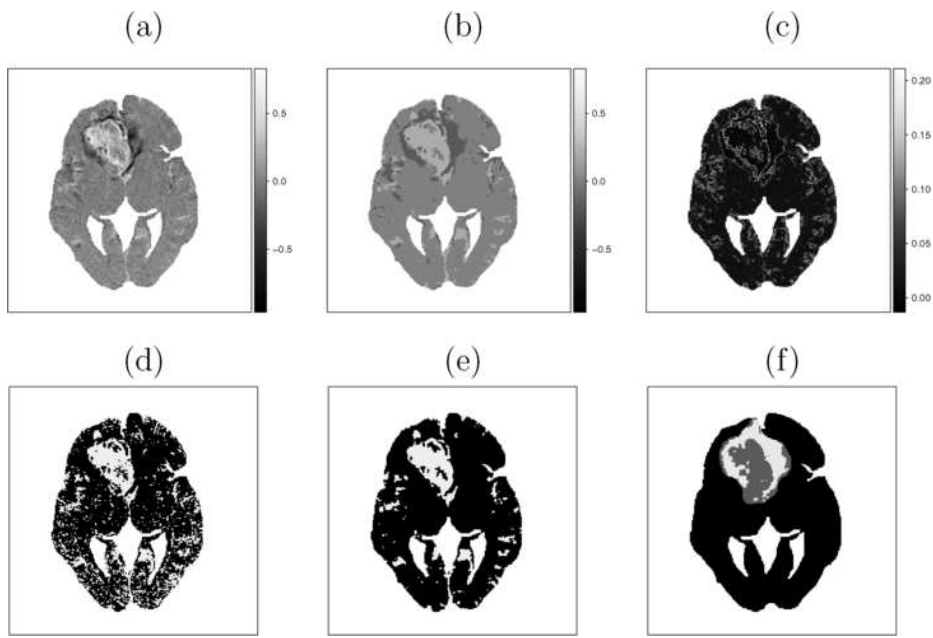

FIG. 2. *Results on patient 1: (*a*) observed change in contrast uptake (light shades standing for large increase); (*b*) expected intensity $\hat{\mu}^{\text{est}}_{z_i}$ and (*c*) standard deviation $\hat{\sigma}^{\text{est}}_{z_i}$; thresholded image (*d*) without and (*e*) with consideration of spatial structure; (*f*) baseline enhanced and nonenhanced tumor regions.*

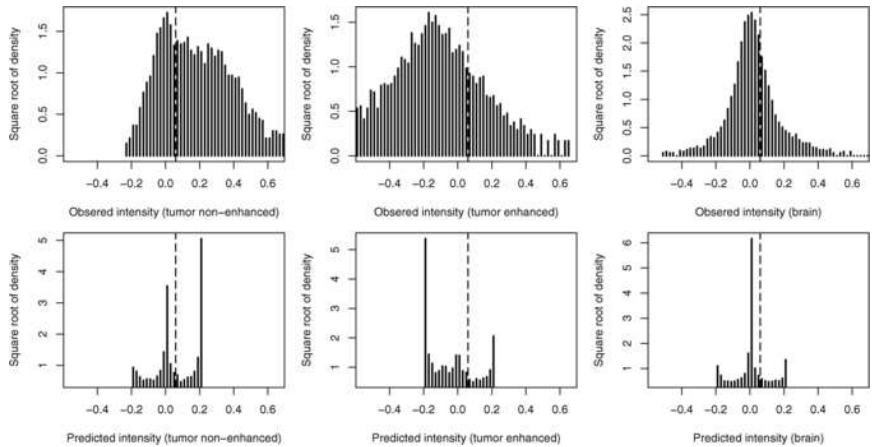

Fig. 3. *Histogram of observed (top) versus estimated (bottom) change in contrast uptake in the healthy tissue (left), initially nonenhanced (middle) and enhanced (right) tumor regions (patient 1).*

The results from patient 2 are similar (Figure 4). Both AIC and BIC favor $M = 4$. The four component mean estimates are $-0.264 \pm 0.004$, $-0.095 \pm 0.001$, $-0.013 \pm 0.0010$ and $0.120 \pm 0.004$ respectively.

As discussed in the Introduction, large increases in contrast uptake are indicative of heavier damage to the BTB/BBB. This suggests that chemotherapeutic agents, in the size range of the contrast medium, may pass the BTB/BBB more easily. Hence, a large increase in the tumor and a small increase (or even decrease) in healthy tissue may suggest the opportunity to deliver these agents more effectively during this window of time. An alternative to comparing the mean change is to define a threshold of change and compare the proportions of healthy and diseased tissue that exceed this threshold. A biologically meaningful threshold of change has not been defined in this exploratory study, but for illustrative purpose, we choose a threshold of 0.06.

Ignoring the spatial structure of the data and thresholding the observed change in uptake (patient 1), regions that lie above the threshold scatter throughout the tumor/brain [41.2% of the tumor and 19.9% of the healthy tissue, Figure 2(d)], much of which, we believe, is attributable to random noise. Our proposed algorithm borrows strength from neighboring pixels, reducing both FPR and FNR and producing a smoother image [Figure 2(e)]. Overall, 39.4% of the tumor exceeds the threshold, while only 7.6% of the healthy tissue exceeds the threshold. For patient 2, we found that 29.8% of the tumor exceeds the threshold as compared to 3.9% of the healthy tissue. The thresholded image is again smoother when we account for spatial correlation [Figure 4(e)] than when we ignore it (27.3% of the tumor and 5.0% of the healthy tissue exceed the chosen threshold, Figure 4(d).

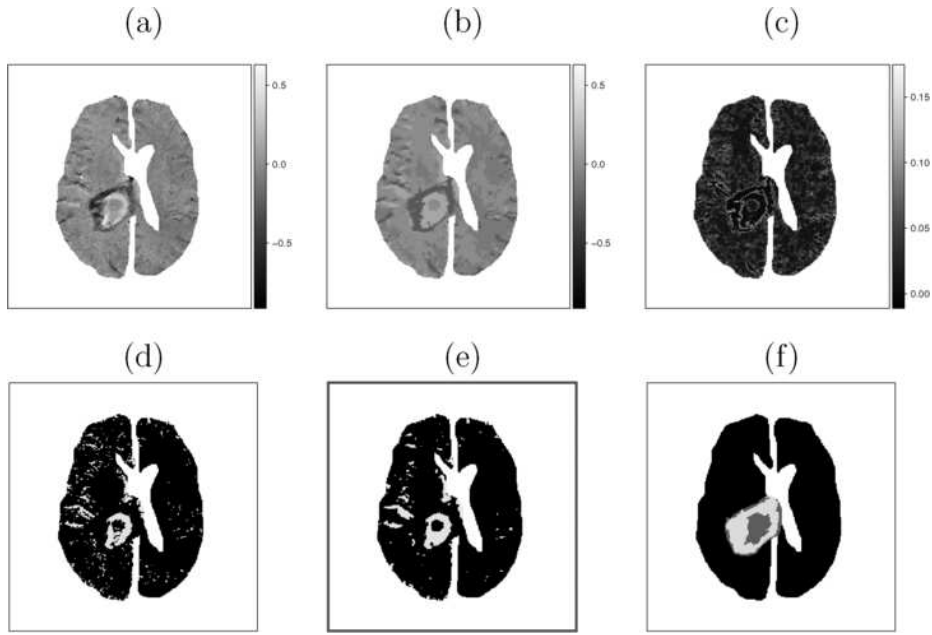

FIG. 4. *Results on patient 2: (*a*) observed change in contrast uptake (light shades standing for large increase); (*b*) expected intensity $\hat{\mu}_{z_i}^{\text{est}}$ and (*c*) standard deviation $\hat{\sigma}_{z_i}^{\text{est}}$; thresholded image (*d*) without and (*e*) with consideration of spatial structure; (*f*) baseline enhanced and nonenhanced tumor regions.*

In the original analysis, Cao et al. (2005) divided the tumor into two regions: One region where the pre-treatment contrast uptake was relatively large and the second region where it was relatively small. These regions typically divided the tumor into a "core" (low initial contrast uptake) surrounded by an annulus (high initial contrast uptake). The biological rationale for this is that the core of the tumor is typically hypoxic (low oxygen content) due to a lack of blood supply, while the annulus of the tumor is rich in blood supply due to angiogenesis (new blood vessel growth, that is typically disorganized in tumors and thus leaky). Hypoxia is known to have a protective effect against damage due to both radiation and chemotherapy. Opening the vascular permeability (increase in contrast uptake) in the tumor core may allow the chemotherapeutic agents to penetrate the BTB and to access tumor cells in the core. If one could predict when this small increase takes place, it may then provide rationale for initiating chemotherapy; thus allowing for more effective control of the tumor core.

To divide the tumor into these two regions, Cao et al. (2005) used one standard deviation above the average contrast uptake in the healthy tissue that received a total dose less than 10 Gy. This number was then used to divide the tumor into initially enhanced (high contrast uptake) and initially nonenhanced regions (low contrast uptake). This criterion is arguably low. Furthermore, a large portion of healthy tissue, that receiving more than 10 Gy, is ignored, as is the spatial correlation inherent in the data. We take a different approach to dividing the tumor. First, we run the proposed algorithm on the baseline contrast uptake image and divide the tumor into initially enhanced and nonenhanced regions based on the 95th percentile of the estimated healthy tissue contrast uptake. As seen in Figures 2(f) and 4(f), the initially enhanced tumor area (in light gray shade) roughly corresponds to an annulus surrounding the nonenhanced area—the core (in medium gray shade). We also note that, in both patients, there is a thin outer annulus of nonenhancing tumor. We suspect that this may be caused by two sources of error. One, the tumor outlines were obtained from a radiation oncologist for radiation planning purpose and therefore may contain a thin margin outside the observed tumor region (to ensure that all the tumor received a uniform dose of radiation). Second, this may be caused by volume averaging (pixels near the edge of the tumor contain both diseased and healthy tissue). Nevertheless, we include this region as part of the initially nonenhanced region in our analysis.

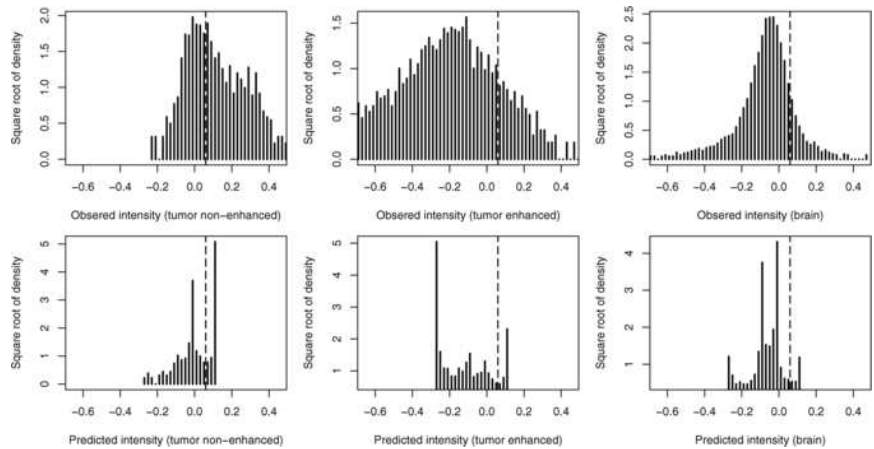

Fig. 5. *Histogram of observed (top) versus estimated (bottom) change in contrast uptake in the healthy tissue (left), initially nonenhanced (middle) and enhanced (right) tumor regions (patient 2).*

TABLE 2
*Observed and estimated percentage of pixels in the initially enhanced and nonenhanced tumor region under various thresholds (90/95/97.5th percentile of healthy tissue contrast uptake before radiotherapy) without and with consideration of the spatial structure*

| Percentile of healthy tissue at baseline | Patient 1 | | | | Patient 2 | | | |
|---|---|---|---|---|---|---|---|---|
| | Enhanced (%) | | Nonenhanced (%) | | Enhanced (%) | | Nonenhanced (%) | |
| | Obs. | Pred. | Obs. | Pred. | Obs. | Pred. | Obs. | Pred. |
| 90th | 17.1 | 16.9 | 64.2 | 60.1 | 11.5 | 14.0 | 53.2 | 55.8 |
| 95th | 14.7 | 14.4 | 62.1 | 59.1 | 10.2 | 12.6 | 51.6 | 54.5 |
| 97.5th | 12.5 | 12.5 | 59.8 | 56.8 | 9.4 | 11.5 | 49.9 | 53.1 |

For patient 1, 59.1% of the initially nonenhanced tumor region has an increase in contrast uptake above the threshold, 0.06, compared to 14.4% in the initially enhanced region (Figure 3). Similarly, for patient 2, 54.5% of the initially nonenhanced tumor region has a change in uptake that exceeds the threshold, compared to 12.6% in the enhanced region (Figure 5).

Admittedly, the choice of 95th percentile for defining the initially enhanced and nonenhanced tumor regions is ad-hoc. We performed a small study to address the sensitivity of our results to this choice. Our tumor division was based on the 95th percentile of the contrast uptake in healthy tissue. We compared our results to those using the 90th and 97.5th percentiles. The percentages of the initially enhanced and nonenhanced regions of the tumors that exceed the various thresholds are given in Table 2. From this table, it is evident that our results are not highly sensitive to the choice of threshold over the range of thresholds studied.

**4. Conclusion.** We have proposed an image smoothing algorithm suitable for qMRI data with edge preservation. Compared to previous work on similar models, we show how to correctly implement the stochastic variation of the EM algorithm. More importantly, we quantify the uncertainty in parameter estimates; previous work targets the hidden labels and point estimation of parameters without attempting to quantify this uncertainty. Furthermore, we focus on the expected change in contrast uptake rather than the hidden labels, which are hard to interpret. The performance of the proposed method is satisfactory in both simulation studies and on real data. The model is rather robust to violations of the model assumption. Although the algorithm works fine under moderate smoothing of the observed image as shown in Section 3.2, the results degrade as the smoothing becomes heavy. We therefore suggest that no additional smoothing of the data be performed after image reconstruction. We also comment that there is no theoretical difficulty to apply the model/algorithm in 3D images. However, the computational complexity will increase dramatically and the simulation time will increase exponentially.

We emphasize that we do not draw any inference on the components. Therefore, we do not think it matters whether or not they have physiological meanings. Nevertheless, components whose means are larger indicate an increase in contrast uptake relative to components whose means are smaller. This differs from an absolute contextual meaning.

In Section 3 we have investigated a pure noise scenario and a few basic edge patterns (horizontal, vertical, diagonal and curved). The performance of the proposed algorithm is satisfactory. As pointed out by a reviewer, we agree that the performance of the algorithm under various scenarios should be explored. Therefore, we generate a more realistic simulation by taking the output from the proposed algorithm on the real data as the "truth" and add on conditionally independent noise (thus, we know the "true" number of components). We then run the proposed algorithm on this simulated data set. The number of components selected by the information criteria are consistent with the "truth."

The EM algorithm is only guaranteed to converge to a local maximum, and the complexity of the data may imply multiple local maxima. There are stochastic variations of the EM algorithm other than the one discussed here, which are of possible interest, such as the Stochastic EM algorithm [Celeux and Diebolt (1985)] and Stochastic Approximation EM [Delyon, Lavielle and Moulines (1999)]. The basic idea of the stochastic variation is to inject random noise into the deterministic update of EM in the hope that the noise will "push" the method away from a local trap and hence lead to a better solution. Some of our preliminary work on the Stochastic EM algorithm suggests similar performance. Although the stochastic variation alleviates some of the trapping, it does not necessarily find all local maxima.

We have focused on model selection, and ignored the uncertainty in this selection. It is possible that a single best model does not exist. Therefore, model averaging is a direction worth exploring. Buckland, Burnham and Augustin (1997) suggested several ad hoc non-Bayesian approaches to account for model uncertainty. The smoothed AIC estimator is later embraced by Burnham and Anderson (2002), Claeskens and Hjort (2003) and Hjort and Claeskens (2003). It essentially constructs a weighted average of parameters of interest across candidate models, where the weight is proportional to the exponent of AIC, that is, $w_m \propto \exp\{-\text{AIC}_m/2\}$. In our application, we have done some preliminary work in this direction. We first compute the estimated intensities $\mu_{z_i}^{\text{est}}$ for each sub-model indexed by $M$. We then apply the above weight to obtain a weighted average across a series of sub-models. However, in our applications, the AIC for the best model is considerably smaller than that for competing models, so model averaging does not make a practical difference. As alternatives to the deterministic nature of the EM algorithm, there are stochastic approaches for similar models using evolutionary algorithms. Destrempes, Mignotte and Angers (2005) extended the earlier work by Francois (2002) using the Exploration/Selection/Estimation procedure, although we have not explored these alternatives.

We are currently working on a parallel Bayesian analysis of our proposed image model where the distribution of the number of hidden labels is estimated via reversible jump MCMC [Green (1995)]. In the Bayesian framework, model averaging is a natural, and often argued, method of parameter estimation, as model uncertainty is accounted for in the estimates.

In this manuscript we have concentrated in the change of contrast uptake from baseline to the three week visit, identified as a key visit. The data actually consist of a baseline image study and five follow-up image studies. Thus, modeling both the spatial and temporal aspects of the study is also of great interest.

X. Zhang
Pfizer Inc.
235 East 42nd Street 685/10/08
New York, New York 10017
USA
E-mail: xiaoxi.zhang@pfizer.com

T. D. Johnson
R. J. A. Little
Department of Biostatistics
University of Michigan
1420 Washington Heights
Ann Arbor, Michigan 48109
USA
E-mail: tdjtdj@umich.edu
rlittle@umich.edu

Y. Cao
Department of Radiation Oncology
University of Michigan School of Medicine
1500 East Medical Center Drive
Ann Arbor, Michigan 48109
USA
E-mail: yuecao@med.umich.edu